\documentclass[twocolumn,prl, superscriptaddress, amsmath, amssymb]{revtex4}

\usepackage{latexsym,epsfig,epstopdf,graphicx}
\usepackage{dcolumn}
\usepackage{bm}

\begin{document}

\title{The dynamics of condensate shells: collective modes and expansion}

\author{C. Lannert}
\affiliation{Wellesley College, Wellesley, MA 02481, USA}

\author{T.-C. Wei}
\affiliation{University of Illinois at Urbana-Champaign, Urbana, IL 61801, USA}

\author{S. Vishveshwara}
\affiliation{University of Illinois at Urbana-Champaign, Urbana, IL 61801, USA}

\date{\today}

\begin{abstract}
We explore the physics of three-dimensional shell-shaped
condensates, relevant to cold atoms in ``bubble traps" and to Mott
insulator-superfluid systems in optical lattices. We study the
ground state of the condensate wavefunction, spherically-symmetric
collective modes, and expansion properties of such a shell using a
combination of analytical and numerical techniques. We find two
breathing-type modes with frequencies that are distinct from that
of the filled spherical condensate. Upon trap release and
subsequent expansion, we find that the system displays
self-interference fringes. We estimate characteristic time
scales, degree of mass accumulation, three-body loss, and kinetic
energy release during expansion for a typical system of
$^{87}$Rb.
\end{abstract}

\maketitle

The cooling and trapping of dilute atoms has recently achieved
unprecedented levels of control and sophistication. With the
advent of optical lattices~\cite{OL}, quasi-one and -two
dimensional trapping potentials~\cite{1&2d}, and mixtures of
different species~\cite{mixtures}, condensates of
bosonic atoms have been created in a plethora of interesting
geometries. Boson mixtures in a particular regime of interactions
can form a layered or core-and-shell structure~\cite{PS}.
Bosons in a three-dimensional optical lattice also display a shell
structure as a result of the confining trap~\cite{bakedAL}. It may
even be possible to confine a dilute atomic
condensate to a spherical shell-shaped region by means of a 
specifically-designed ``bubble trap"~\cite{ZG}. Towards an understanding of these, and
other systems where superfluid order exists in ``hollow"
geometries, we consider the physics of a condensate whose shape is
a three-dimensional spherical shell. We identify key features in
the condensate collective modes and expansion upon trap release
that would distinguish such shell-shaped condensates from the
more common filled cases. Moreover, we find that expansion
properties have distinct similarities with the ``Bose-nova"
experiments~\cite{BN}.

As a potentially striking application, our study of condensate
dynamics ought to provide a novel probe for the predicted
co-existence of shell-shaped superfluid and Mott insulator regions
in optical lattices~\cite{bakedAL}. At the mean field level, and at
scales much larger than the optical lattice spacing, the physics of the superfluid regions should be
well-described by a smooth effective potential representing the
Mott regions. In fact, our analyses ought to be relevant to any
mean-field description which yields shell-shaped
condensate confinement, as for instance in the case of boson
mixtures where repulsive interactions of one species exclude
another species from occupying the center of the trap. 

We therefore consider the ground state wavefunction for the
condensate assuming an effective three dimensional confining
potential of the form $V(\mathbf{r}) = ({1}/{2}) m \omega_0^2
(r-r_0)^2 $ (where $m$ is the atomic mass). The ground state
wavefunction of the condensate obeys the time-independent
Gross-Pitaevskii (GP) equation:
\begin{equation}
\left[ -\frac{\hbar^2}{2m} \nabla^2 + V(\mathbf{r}) + {g} \left| \psi (\mathbf{r})\right|^2 \right] \psi (\mathbf{r})  \equiv \frac{\delta \cal H}{\delta \psi^*} =  \mu \psi (\mathbf{r}),
\label{GP}
\end{equation}
where $g$ is a measure of the repulsive interactions between the
atoms ($g=4\pi\hbar^2 a/m$ with $a$ the $s$-wave scattering
length) and $\mu$ is the chemical potential, set by the
normalization condition: $\int  \left| \psi(\mathbf{r}) \right|^2
d^3 r = N $. The density of condensed atoms is given by
$n(\mathbf{r}) = |\psi(\mathbf{r})|^2 $.

For suitably large $N$, the kinetic energy is very small compared
to the potential energies and the gradient term may be neglected
(the Thomas-Fermi approximation)~\cite{Baym96}, giving an
approximate ground state wavefunction $\psi_{TF} (\mathbf{r}) =
\sqrt{[\mu-V(\mathbf{r})]/g}$ in the regions where $\mu \geq
V(\mathbf{r})$ and zero elsewhere. In this approximation, the
condensate occupies a spherically-symmetric shell centered at
$r_0$ with half-width $r_1 = \sqrt{2\mu/(m\omega_0^2)}$. 
 In the limit of a thin shell ($r_1 \ll r_0$), $r_1$ is found to be
$[3gN/(8\pi m \omega_0^2 r_0^2)]^{1/3}$. Therefore, for the
Thomas-Fermi wavefunction, the condition for a thin shell can be written
\begin{displaymath}
\Gamma_{\rm ts}\equiv \frac{r_1}{r_0} = \left(\frac{3gN}{8\pi m \omega_0^2r_0^5}\right)^{1/3}=\left(\frac{3Na_{\mbox{\scriptsize osc}}a}{2r_0^5}\right)^{1/3}\ll 1,
\end{displaymath}
where we have introduced the oscillator length $a_{\mbox{\scriptsize osc}} \equiv \sqrt{\hbar/(m\omega_0)}$. In order to obtain analytic results, we will often work within this ``thin shell" limit. 
While applicable to typical conditions in bubble traps and optical lattice systems, we also expect an analysis of thin shells to capture the salient features of thicker shells. Notable expected differences will be discussed in some cases.

While we are able to go beyond the Thomas-Fermi
approximation using numerical techniques, it is worth estimating its regime of
validity. Using a radially symmetric Gaussian wavefunction centered
at radius $R_0$ and with characteristic width $R_1$, we find the
ratio of the zero-point kinetic energy to the potential energy of
interaction to scale as $R_0^2/(NaR_1)$. (The analogous ratio for a
condensate in a harmonic trap centered at $r=0$ scales as
$R/(Na)$ (see, e.g.,~\cite{PS})). Hence, for the shell (i.e., taking
$R_0$ to be $r_0$ and $R_1$ to be $r_1$),
 the Thomas-Fermi approximation can be expected to be valid when $
\Gamma_{\rm TF}\equiv r_0^2/(Na\,r_1)= \left( [{2r_0^8}]/[{3N^4
a_{\mbox{\scriptsize osc}}^4 a^4}] \right)^{1/3} \ll 1$.

For a trap with $\omega_0= 2\pi \times 20$~Hz and $r_0=20
a_{\mbox{\scriptsize osc}}$ and a cloud of $N=10^6$ atoms of
${}^{87}$Rb in the $|F=1,m_F=-1\rangle$ or $|F=2,m_F=2\rangle$
state (for which $a\approx5.45$~nm~\cite{GreinerThesis}), we have $\Gamma_{ts}\ = {r_1}/{r_0} \approx 0.08$ and
$\Gamma_{TF} \approx 0.09$. In excellent numerical agreement, using
the imaginary time technique of Chiofalo \emph{et
al.}~\cite{ChiofaloIm} and the parameters given above, we find the
ratio of the kinetic energy to the total energy in the ground
state to be $K/E_{tot} \simeq 0.093$.

Towards obtaining the lowest-energy collective modes of a thin
spherical shell of superfluid, we consider a trial wavefunction
for the shell condensate of the form:
\begin{equation}
\psi_{trial} [R_0, R_1] = {\cal A}\frac{\sqrt{N}}{R_0 \sqrt{R_1}} {\cal F} \left( \frac{r-R_0}{R_1} \right) e^{i \phi(\mathbf{r})} ,
\label{varwf}
\end{equation}
where ${\cal A}$ is a dimensionless normalization constant (in the
thin shell limit) and ${\cal F}$ is a smooth real function that is
negligibly small for $|r-R_0| \gg R_1$, (for instance a Gaussian:
${\cal F} (x) = e^{-x^2/2}$). We note that except for the
discontinuity at the condensate boundary, the Thomas-Fermi
wavefunction is in the class of wavefunctions described by
Eq.(\ref{varwf}). The function $\phi(\mathbf{r})$
has the usual relation to the velocity of the condensate:
$\mathbf{v} = {\hbar} \nabla \phi/m$. $R_0$ and $R_1$ are
variational parameters corresponding to the average radius and
characteristic width of the cloud, respectively.  A description of
the cloud in terms of the parameters $R_0$ and $R_1$ is expected
to capture the salient features of the isotropic, $\ell =0$,
collective modes of the superfluid shell.

\emph{Collective modes.} Starting with the variational
wavefunction in Eq.(\ref{varwf}) and an appropriate velocity
field, $\mathbf{v}(\mathbf{r})$, standard methods yield equations
of motion for the collective coordinates $R_0$ and $R_1$ (see, e.g.,~\cite{PS}). We start by noting that the energy of the condensate
for a wavefunction of the form Eq.(\ref{varwf}) can be written
as:
\begin{equation}
{\cal H}[R_0,R_1,\phi] =  \frac{m}{2}\int n(\mathbf{r}) |\mathbf{v}(\mathbf{r})|^2 d^3 r + {\cal U}_{eff} ,
\label{VarEn}
\end{equation}
 where ${\cal U}_{eff}$ is equal to the energy of the cloud if the phase $\phi$ does not vary in space and acts as an effective potential for the parameters $R_0$ and $R_1$. It can be written in a physically transparent form as a sum of contributions from zero-point (or confinement) energy, potential energy from the trap, and interaction energy: ${\cal U}_{eff}[R_0,R_1] = E_{zp} + E_{tr} + E_{int}$.
For the trial wavefunction in Eq.(\ref{varwf}), in
the thin shell limit ($R_1 \ll R_0$), we find
 \begin{eqnarray*}
E_{zp} &=& \frac{\hbar^2}{2m} \int \left( \frac{d |\psi|}{dr}
\right)^2 d^3 r \approx \frac{c_{zp}}{R_1^2} ,\\
E_{tr} &=& \int V(\mathbf{r}) |\psi|^2 d^3 r \approx  \frac{Nm}{2} \omega_0^2 \left[ c_{tr} R_1^2 +  (R_0-r_0)^2 \right] ,\\
E_{int} &=& \frac{g}{2} \int  |\psi|^4 d^3 r \approx \frac{c_{int}}{R_0^2 R_1},
\end{eqnarray*}
to lowest nonvanishing order in $R_1/R_0$, where $c_{zp}$,
$c_{tr}$, and $c_{int}$ are independent of $R_0$ and $R_1$ and are
determined by the form of the function ${\cal F}$ in
Eq.(\ref{varwf}).

The variational energy, Eq.(\ref{VarEn}) can be used to find two, low-energy,
collective excitations of the superfluid shell: one in which the width,
$R_1$, oscillates around its equilibrium value (the ``accordion
mode") and another in which the average radius of the cloud,
$R_0$, oscillates around its equilibrium value, $r_0$ (the
``balloon mode"). For simplicity, we assume that in the accordion
mode, the mean radius of the shell stays fixed at $r_0$ while the
width oscillates and that in the balloon mode the width, $R_1$,
remains fixed while the mean radius oscillates. While it is clear
that any exact solution will couple changes in the width to
changes in the mean radius, the ``decoupled" oscillations we
consider here can be expected to illuminate the correct low-energy
physics. Indeed, in the thin shell limit, we find that
oscillations in $R_1$ do not affect the mean radius $R_0$ and that
oscillations in $R_0$ affect the width, $R_1$, at a negligible
level for small oscillations about equilibrium (smaller by a
factor $[R_0-r_0]/r_0$).

\emph{The Balloon Mode.} For this mode, we consider a velocity
field of the form $\mathbf{v}_b = \beta \hat{r}$, where $\beta$ is
a variational parameter, or equivalently, $\phi(\mathbf{r}) =
\beta m r/\hbar$. As a lowest-order approximation, we hold $R_1$
fixed at its equilibrium value, $r_1$, and only allow $R_0$ to
vary in time. By constructing a Lagrangian for the parameters
$R_0$ and $\beta$, we find $\beta = \dot{R}_0 \Rightarrow
\mathbf{v}_{b} = \dot{R}_0 \hat{r}$ and an equation of motion for
$R_0$:
\begin{equation}
mN \ddot{R}_0 = -\frac{\partial {\cal U}_{eff}}{\partial R_0} = \frac{2c_{int}}{R_0^3 r_1} - Nm \omega_0^2 (R_0-r_0). \label{R0}
\end{equation}
In the thin shell limit, Eq.(\ref{R0}) yields $R_0^{eq} = r_0$ and a frequency of small oscillations around this equilibrium value:
\begin{equation}
\omega_{b} \simeq \omega_0 + {\cal O}(r_1^2/R_0^2).\label{omega-b}
\end{equation}
We note that the thin-shell approximation imposes a constraint on
the amplitude of oscillations in $R_0$: $R^{min}_0 \gg r_1$.

\emph{The Accordion Mode.} For this mode, we consider a velocity field for the condensate of the form
$\mathbf{v}_a = \beta (r-r_0) \hat{r}$ (equivalently: $\phi (\mathbf{r}) = \beta m(r-r_0)^2/(2\hbar) $) and allow $R_1$ to vary in time while holding $R_0$ fixed at its equilibrium value, $r_0$.  Following the same procedure as for the balloon mode, we find $\beta = {\dot{R}_1}/{R_1} \Rightarrow \mathbf{v}_a =  \hat{r}(r-r_0){\dot{R_1}}/{R_1}$ and an equation of motion for $R_1$:
\begin{equation}
m_{eff}^a \ddot{R}_1 = -\frac{\partial {\cal U}_{eff}}{\partial R_1} = \frac{2c_{zp}}{R_1^3} + \frac{c_{int}}{R_1^2 r_0^2} - m_{eff}^a \omega_0^2 R_1,
\end{equation}
with $m_{eff}^a \equiv 4\pi m{\cal A}^2 N \int_{-\infty}^{\infty} q^2 {\cal F}^2(q) dq$. Using the fact that $\partial {\cal U}_{eff} / \partial R_1 = 0$ at equilibrium, the frequency of small oscillations of $R_1$ about its equilibrium value can be written:
\begin{equation}
\omega_a = \omega_0 \sqrt{ 4- \frac{E_{int}}{2 E_{tr}}}.
\label{omega-a}
\end{equation}
The thin-shell approximation imposes a constraint on the amplitude
of oscillations in $R_1$: $R_1^{max} \ll r_0$. In the limit of
weak interactions, $E_{int} \ll E_{tr}$, Eq.(\ref{omega-a}) reduces to $\omega_a = 2
\omega_0$. Thus, in the weak interaction limit, the frequency of this mode is equivalent to that of the
breathing mode of a spherical condensate cloud. In the limit of strong interactions, $E_{int}
\gg E_{zp}$, we have $E_{int} \simeq 2 E_{tr} $ and find $\omega_a
= \sqrt{3} \omega_0$. This result should be compared with the
strong interaction limit of the spherical breathing mode: $\omega_{br}
= \sqrt{5} \omega_0$.

We note that the modes in the thin-shell limit described by
Eq.(\ref{omega-b}) and Eq.(\ref{omega-a}) have a structure
identical to that of a one-dimensional condensate, corresponding
to its 1-d sloshing and breathing modes, respectively.
For thicker
shells, we expect corrections to our results that couple the $R_1$
and $R_0$ degrees of freedom. In fact, in the limit that $R_0
\rightarrow 0$ we expect the balloon and accordion modes to tend
to the breathing mode and next radially symmetric mode ($n=2$) of
a filled spherical condensate with $R_0=0$.

\emph{Expansion.} The dynamics of the spherical shell upon release
of the trapping potential has noteworthy features absent in the
case of the filled sphere. Upon release, the initial confinement
of the condensate causes the outer edge to expand outwards and the
inner edge to collapse inwards. As a result, the system can
potentially exhibit accumulation of mass at the center, and the
condensate can interfere with itself when diametrically opposite
regions come together.

The timescale of expansion can be estimated within the thin-shell
approximation (where the dynamics of the width, $R_1$, do not
affect the mean radius, $R_0$, which we approximate as fixed at
$r_0$). Taking the function ${\cal F}$ in Eq.(\ref{varwf}) to be a Gaussian and
evaluating the different energy contributions as before, we
find that
energy conservation between the instant the trap is
switched off and later times $t$ gives the relationship
\begin{displaymath}
\frac{\hbar^2}{m R_1^2(0)} +
\frac{gN}{{(2\pi)}^{3/2} R_1(0)r_0^2} =
m\dot{R}_1^2 + \frac{\hbar^2}{m R_1^2} +
\frac{gN}{{(2\pi)}^{3/2} R_1r_0^2}, 
\end{displaymath}
where the time argument of $R_1(t)$ on the r.h.s is suppressed,
and $R_1(0)$ is the characteristic width of the condensate
shell before expansion.
Assuming the initial energy is dominated by the interaction energy,
we find that, on the time scale for which the shell expands enough to
become dilute but not enough to reach the center,
$R_1^2(t)/R_1^2(0)\approx 2\,\omega_0^2\,t^2$, in contrast to a
filled sphere for which $R_1^2(t)/R_1^2(0) \approx
({2}/{3})\omega_0^2\,t^2$~\cite{PS}. Hence, for typical parameters used in
this paper, a shell of initial thickness $5\,\mu{\rm m}$ and radius
$50~\mu$m should expand to a thickness of $R_1(t)=20~\mu$m on a
time scale of around 20~ms, which is amenable to experimental detection.

\begin{figure}[t]
\vspace{-0.2cm}
\centerline{\psfig{figure=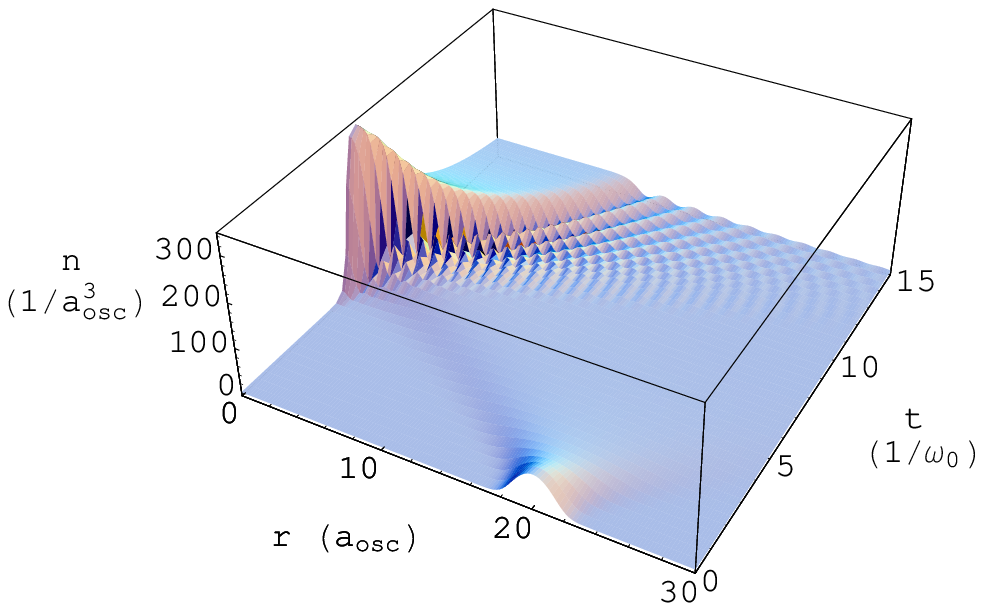,width=7cm,angle=0}}
\vspace{-0.3cm}
\centerline{\psfig{figure=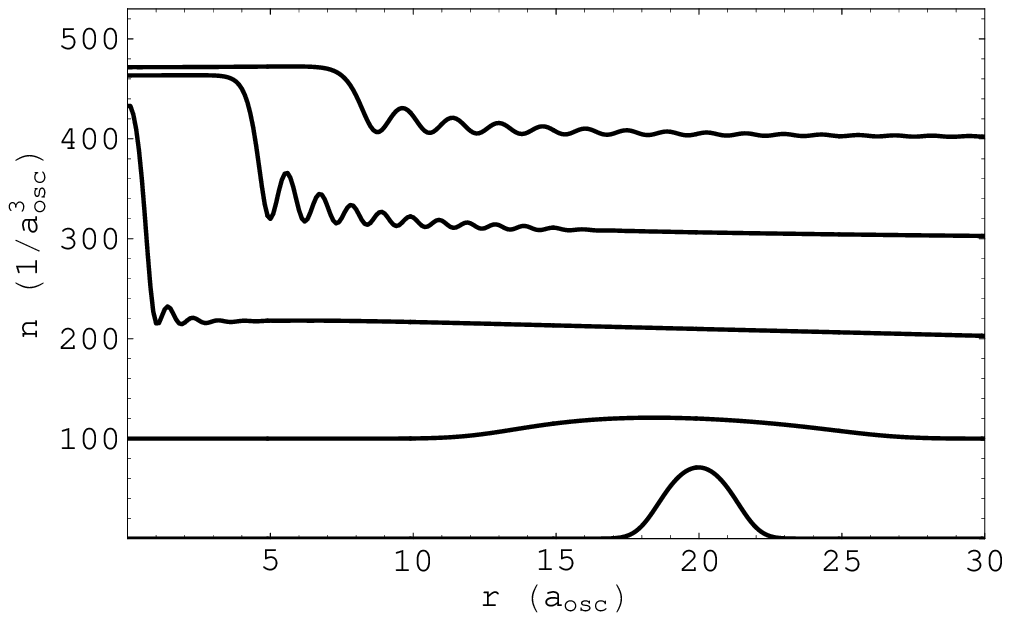,width=7cm,angle=0}}
\caption{Expansion of a thin spherical shell. 
(a) Evolution of the density as a function of radial
position and time. (b) Snapshots of the density profile at times
$t=0,3,6,9,12$ (from the lowest curve, subsequent curves are shifted
up by 100 for clarity).\label{fig:Expand} }
\end{figure}

To quantify the physics of the expanding shell,
we performed a numerical time evolution of the initial condensate wavefunction (given by the numerically-obtained result discussed earlier) after release from the trap, including interactions. This expansion process was obtained
using the real-time synchronous Visscher method~\cite{ChiofaloRe} to
integrate the time-dependent GP equation: $i \,d\psi/dt = \delta {\cal H}/\delta \psi^*$. The results are shown in Fig.~\ref{fig:Expand}.

Two general features of the expansion deserve discussion.
First, mass accumulation in the center can result in a density
greater (but not necessarily much greater) than the initial density.
We note that
repulsive interactions between the atoms prevent the density from
becoming as large as it would in the non-interacting case. Secondly,
interference fringes appear after the inner radius of the cloud has
reached the origin, demonstrating self-interference of the
condensate. For two Gaussian condensates initially separated by a
distance $D$, the fringe size at
long times is given by $\delta_r=2\pi \hbar t /(m D)$~\cite{PS} in the absence of interactions.
The free expansion of an initially thin Gaussian
shell is straightforward to calculate and we find that the fringe spacing at long times is identical to that of two Gaussian condensates, but with the initial separation, $D$, replaced by the initial diameter of the shell, $2\,r_0$.
This implies $\delta_r = \pi\,t\,\omega_0 \,a^2_{\mbox{\scriptsize osc}}/r_0 $, which for $r_0=20\, a_{\mbox{\scriptsize osc}}$ gives
a fringe spacing at time $t\,\omega_0=10$ of $\delta_r \simeq 1.6\, a_{\mbox{\scriptsize osc}}$. This compares to an average fringe size observed in our numerics of
about $1.2\, a_{\mbox{\scriptsize osc}}$. The difference in precise values
presumably results from the effects of interparticle interactions and the non-Gaussian shape of the initial wavefunction.

The dynamics of the shell upon trap release has distinct parallels
with the Bose-nova collapse of a condensate when the interatomic interactions are switched
from repulsive to attractive. For the shell, the
initial implosion is caused by the quantum pressure of the
condensate forcing itself to fill the low-density region at the
center. Similar to the Bose-nova case, mass buildup near the origin
 is followed by a relaxation and expansion on timescales
comparable to the trap frequency. An important question, given these
parallels, is whether three-body recombination and subsequent
``loss" of atoms is appreciable in the case of the released shell
(as it is in the case of the Bose-nova collapse~\cite{BN}). These
losses are described by the equation $dn/dt = - K_3 n^3$ where $K_3$
is the three-body loss rate~\cite{PS}. Concentrating on the
time in our numerics with the largest density (at $t \approx
6.5/\omega_0$), we estimate that the density in the
central plateau is $n \simeq 350/a^3_{\mbox{\scriptsize osc}}$
over a radius of about $a_{\mbox{\scriptsize osc}}$.
Assuming that this density persists for the entire time between
snapshots, $\delta t \approx 3/\omega_0$, and taking $K_3 = 4.0
\times 10^{-30}~\mbox{cm}^6/\mbox{s}$~\cite{PS}, we find an upper
bound on the number of particles lost in this region during this
time slice of $\Delta N \simeq 0.09$, making
three-body recombination negligible for the case considered here.

The effects of mass accumulation would be enhanced if only the
inner edge of the trap were removed, suppressing the outward
expansion of the condensate. This more dramatic case was
 considered by Zobay and Garraway~\cite{ZG}, in which
they modeled an initially shell-shaped trap quickly switched into
a harmonic trap. Similar features of mass accumulation and
self-interference fringes were found in the case of this bubble
trap. For a shell with the parameters given in this paper, we
can estimate the timescale for collapse and the kinetic energy
gain in this scenario by considering a small cavity of radius
$R_2$ at the center of a condensate. In the Thomas-Fermi
approximation, the dynamics of such a cavity can be mapped to the
standard hydrodynamics problem of a collapsing bubble in a fluid
governed by the equation~\cite{lamb}
\begin{equation}
\frac{p-p_0}{\rho}=\frac{R^2_2\ddot{R_2}+2R_2\dot{R_2}^2}{r}-\frac{R_2^4\dot{R}^2}{2r^4}.
\label{hydrodyn}
\end{equation}
Here, $\rho=mn$ is the condensate mass density, $p$ is the
pressure at radius $r$ and $p_0={n^2g}/{2}$ is the pressure
far from the cavity. Integrating Eq.(\ref{hydrodyn}) at the edge
of the bubble ($r=R_2$) and making the substitution
$R_2(t)=R_2(0)x^{1/3}$ gives the time for complete collapse in
terms of the initial radius, $R_2(0)$:
$t_f\approx0.915R_2(0)\sqrt{\rho/p_0}$. The kinetic energy gained
by the particles upon reaching the center is given by $E_{KE}=4\pi
/3 p_0 R_2^3(0)$. For a cavity of radius $40~\mu$m and quantum
pressure of magnitude $p_0=1\times10^{-14}~\mbox{erg/cm}^3$, the
collapse time is on the order of 100~ms and the kinetic energy
gained per particle is on the order of 1~nK. The small cavity
treatment suggests that the collapse of the inner radius can be
accompanied by a measurable amount of kinetic energy
gain and mass accumulation.

In conclusion, motivated by possible new trapping potentials for
Bose-Einstein condensed atoms, as well as inhomogeneous phases of
dilute ultracold bosons in optical lattices or in mixtures, we
have explored the collective modes and expansion dynamics of a
superfluid confined to a spherical shell. The two breathing modes
we find offer a possible way of distinguishing a filled spherical
condensate from a one which is hollow or surrounding some other
phase. The expansion properties of the shell after release from
the trap are found to have some notable similarities with
Bose-nova physics, particularly mass accumulation and
self-interference.

We would like to acknowledge helpful discussions with R. Barankov,
G. Baym, M.-L. Chiofalo, A.J. Leggett, H.-H. Lin, C.-Y.
Mou, D. Pekker, K. Son, M. Stone, D. Wang,  and to thank B. DeMarco for
valuable insights. C.L. and S.V. are grateful for the hospitality of
the Aspen Center for Physics, where much of this work was carried
out. This work was supported in part by NSF Grant No. NSF EIA01-21568.

\end{document}